\documentclass[manuscript]{aastex}
\usepackage{natbib}
\usepackage{graphicx}
\usepackage{txfonts}

\shorttitle{Triggering viscosity and spectral states}
\shortauthors{S. Mondal, S. K. Chakrabarti, S. Nagarkoti and Patricia Ar\'evalo}

\begin{document}

\title{Possible range of viscosity parameter to trigger black hole candidates to exhibit different states of outbursts}

\author{Santanu Mondal~\altaffilmark{1,2}, Sandip K. Chakrabarti\altaffilmark{3,2}, Shreeram Nagarkoti\altaffilmark{2}, and Patricia Ar\'evalo\altaffilmark{1}} 
\altaffiltext{1}{Instituto de F\'isica y Astronom\'ia, Facultad de Ciencias, Universidad de Valpara\'iso, Gran Bretana N 1111, Playa Ancha, Valpara\'iso, Chile}
\altaffiltext{2}{Indian Center for Space Physics, 43 Chalantika, Garia Station Rd., Kolkata, 700084, India}
\altaffiltext{3}{S. N. Bose National Centre for Basic Sciences, Salt Lake, Kolkata, 700098, India.}

\email{santanuicsp@gmail.com; chakraba@bose.res.in; srnagarkoti@csp.res.in; patricia.arevalo@uv.cl}

\begin{abstract}

In a two component advective flow around a compact object, a high viscosity Keplerian disk is
flanked by a low angular momentum and low viscosity flow which forms a centrifugal
pressure supported shock wave close to the black hole.
The post-shock region which behaves as a Compton cloud becomes progressively smaller
during the outburst as the spectra change from the hard state to the soft state
in order to satisfy Rankine-Hugoniot relation in presence of cooling.
The resonance oscillation of the shock wave which causes low frequency quasi-periodic oscillations (QPOs)
also allows us to obtain the shock location from each observed QPO frequency.
Applying the theory of transonic flow with Compton cooling and viscosity, we obtain the viscosity
parameter $\alpha_{SK}$ required for the shock to form at those places in the low-Keplerian component.
When we compare the evolution of $\alpha_{SK}$ for each outburst, we arrive at a major conclusion:
In each source, the advective flow component typically requires exactly a similar
value of $\alpha_{SK}$ when transiting from one spectral state to another (e.g., from hard state
to soft state through intermediate states and the other way around in the declining phase).
Most importantly, these $\alpha_{SK}$ values in the
low-angular momentum advective component are fully self-consistent in the sense that they remain below critical
value $\alpha_{cr}$ required to form a Keplerian disk. For a further consistency check, we compute the
$\alpha_K$ of the Keplerian component, and find that in each of the objects, $\alpha_{SK}$ $<$ $\alpha_{cr}$ $<$ $\alpha_K$.

\end{abstract}


\section{Introduction}
Accreting material from the companion forms a differentially rotating disk (Lynden-Bell \& Pringle, 1974), 
known as an accretion disk around a black hole due to inward transport of mass. The matter from the Companion
passing through the Roche Lobe located at $r=r_{RL}$ will be Keplerian in nature but it requires a large viscosity 
to transport the angular momentum of ${\dot m_d} (l_{RL}-l_{ms})$ per second, where ${\dot m}_d$, $l_{RL}$ and $l_{ms}$
denote the accretion rate in the Keplerian component (in Eddington rate unit), specific angular momenta at the Roche Lobe 
(generally at hundreds of thousands of Schwarzschild radii away) and the marginally stable orbit 
(at $\sim 3$ Schwarzschild radii) for a non-rotating black hole). In fact, almost one hundred
percent of angular momentum must be transported away. In case such a viscous process
is not available, low angular (sub-Keplerian, or advective) `halo' matter from, say, winds of the companion can still
rapidly accrete with a different rate of ${\dot m}_h$. This does not require
high viscosity at all and due to transonicity condition at the inner sonic point (Chakrabarti 1996)
the flow will produce centrifugal force supported standing (Chakrabarti 1989a, hereafter C89a; Igumenshchev et al. 1998; 
Lu \& Yuan 1998; Le \& Becker 2005) or oscillating (Molteni et al. 1996, hereafter MSC96) shock. 
Two nonaxisymetric shock may form near the black holes,
when angular momentum is misaligned with the black hole spin axis (Fragile et al. 2007; Generozov et al. 2014).
In the language of Shakura \& Sunyaev (1973), 
where the viscous stress is replaced by $-\alpha P$ ($P$ is isotropic pressure inside the 
disk and  $\alpha$ is a dimensionless quantity $< 1$) $\alpha_{SK}$ could be very 
low while $\alpha_K$ could be very high. This is more so because the pressure 
inside an efficiently radiating Keplerian disk could be negligible as compared to that in the hot halo 
component. Indeed, recent simulations of Giri et al. (2015, and references therein) 
clearly show the segregation of the two components in a flow where $\alpha$ monotonically decreases 
in the vertical direction. Here, two components of the flow suggests one is Keplerian component, which is mainly
high angular momentum and high viscosity flow at the equatorial plane and the second component is relatively low angular 
momentum and low viscous flow. We denote the viscosity of the Keplerian flow as $\alpha_K$ and sub-Keplerian as $\alpha_{SK}$ 
respectively throughout the paper. The sub-Keplerian component of the flow moves faster due to low viscosity and forms corona, 
which up scatters the soft photons of the Keplerian disk, which moves relatively slower due to high viscosity.
 
In the dwarf novae outbursts repeated events are thought to be due to thermal-viscosity instability (Meyer \& Meyer-Hofmeister 1984; 
Smak 1984; Cannizzo, Chen \& Livio, 1995; Cannizzo et al. 2010). 
An outburst is assumed to be triggered when the accretion rate in the Keplerian component goes up. This could
in turn be due to enhancement of convective viscosity (Bisnovatyi-Kogan \& Blinnikov 1977) 
at farther out on the disk when matter piles up. This process may also increase 
rates of the sub-Keplerian flow. Indeed, analysis of a number of
objects using the two component solution (Debnath et al. 2014; Mondal et al. 2014; Jana et al. 2016; Molla et al. 2016)
indicate that both the accretion rates increase, although the sub-Keplerian flow reaches 
it's peak accretion rate quicker than the time when 
the Keplerian disk reached its peak accretion rate because the former moves at the free-fall (being of low angular momentum) time scale, while
the latter moves in the viscous time scale. These two time scales are clearly seen in all low mass X-ray binaries where the Keplerian disk is large. The time lag is due to viscous delay between the Keplerian and sub-Keplerian components of the flows with different viscosity parameter values (Mondal et al. 2015; Nagarkoti \& Chakrabarti 2016ab, hereafter NC16a, b) and the analysis provides us with 
$\alpha_K$ in each case (NC16b). This time lag found in low mass X-ray sources
is interpreted as the signature of the two component flow (Smith et al. 2001, 2002) with a large Keplerian disk.
It is well known (Chakrabarti, 1996 and references therein) that at a critical viscosity parameter, the topology of a transonic solution changes. For $\alpha>
\alpha_{cr}$ the matter forms a Keplerian disk, very similar to a standard disk if the flow is optically 
thick. However, unlike a standard disk, this flow passes through the inner sonic point before entering 
into the black hole and does not have any truncation radius at marginally stable orbit. For $\alpha<\alpha_{SK}$ the flow may pass through 
two sonic points and form a shock if the Rankine-Hugoniot (RH) relation is satisfied 
(Chakrabarti 1996; Mondal \& Chakrabarti 2013). The shock location is uniquely
determined by the sub-Keplerian flow parameters including its accretion rates and $\alpha_{SK}$.  

Outbursting sources have another interesting property. Typically, the spectral states change from hard states (HS) 
to the soft states (SS) through hard intermediate state (HIMS) and soft intermediate state (SIMS)  
(Belloni et al. 2005; McClintok \& Remillard 2006, for a review; Nandi et al. 2012; Debnath et al. 2013).
The opposite is true in the declining phase of the outburst. When the spectra are fitted
(Debnath et al. 2014; Mondal et al. 2014; Jana et al. 2016; Molla et al. 2016) with two component advective flow 
(TCAF) solution of Chakrabarti \& Titarchuk (1995, hereafter CT95), 
it is easily seen that there is a general trend of increasing both the accretion 
rates and decreasing the shock locations and the shock compression ratios
($R$, the ratio of the pre-shock velocity and the post-shock velocity) in the rising phase 
and the opposite sequence is present in the declining phase. Since the post-shock region
behaves as the Comptonizing cloud, the cooling time scale also varies with shock location and 
the compression ratio. In Chakrabarti et al. (2015, and references therein) it has been established that when the
cooling time scale and the compressional heating, i.e., infall time scale roughly match, the
shocks oscillate. This has also been established by detailed numerical simulations (Garain et al. 2014).
This shock oscillation causes modulation of predominantly Comptonized photons which
is manifested at the quasi-periodic oscillations or QPOs. In the 
so-called propagatory-shock oscillation model (POS, Chakrabarti et al. 2005), the shock location 
can be computed from the QPO frequency in each day.

In a transonic flow solution which includes cooling and viscosity, Mondal et al. (2015) computed 
the $\alpha_{SK}$ value for the outbursting candidate H~1743-322 using observed QPOs as a reference.
Cooling reduces the post-shock pressure and thus the shock moves in to satisfy the RH condition.
Recently, Nagarkoti \& Chakrabarti (2016a) calculated the upper limit of $\alpha_{SK}$ of any transonic flow which allows shock formation in the advective component.
This ranges from $0.05-0.1$ (depending on the other flow parameters) using standing shock condition below which 
the shocks may form in the flow. Titarchuk \& Osherovich (2000), while estimating the Reynolds number 
near a transition layer found that the barrier oscillates to produce QPOs only if $\alpha_K$ (alpha parameter in
a Keplerian disk) is less than $0.3$. King et al. (2007) found that the typical range of 
$\alpha_K$ is between $0.1$ and $0.4$. These numbers are about an order of 
magnitude higher than the results of numerical simulation works. 
Magnetorotational instability models (Hawley \& Balbus 1992; Brandenburg et al. 1995; Hawley
et al. 1995) produced the most promising numerical results of $\alpha_{SK} \sim 0.01$ if no net vertical magnetic 
flux is imposed from outside (Sano et al. 2004; Pessah et al. 2007). 
Hirose et al. (2014) showed that convection enhances MRI turbulent stress during the outburst 
time and increase $\alpha_{SK}$ from $\sim 0.03$ to above $0.1$. 

In the literature the models used for spectral states are generally different from those present to
explain the QPOs (e.g., Galeev, Rosner \& Viana 1979; Haardt \& Maraschi 1993; Zdziarski et al. 2003;
Nowak \& Wagoner 1991; Titarchuk et al. 1998; Shirakawa \& Lai 2002; Kato 2008; Ar\'evalo \& Uttley 2006). 
In contrast, our unifying view  using the properties of transonic flows appears to be more appealing as it does not 
change on a case by case basis.

Moreover, unlike our proposal, models discussed in the literature do not arise out of 
governing hydrodynamic equations. For instance, in a truncated disk model (Esin et al. 1997),
the truncation radius is arbitrary, while in our solution it is the location of the 
standing shock obtained from RH conditions. When QPOs are computed from 
truncation radius, some axisymmetric blobs (Done \& Gierli\'nski 2007 and references therein) 
are assumed to be always present in the disk at that radius (whole origin and stability are not discussed), 
while in our case, the oscillation is due to resonance between cooling and heating time scales. 
Our completely self-consistent picture also explains why mostly the Comptonized photons participate in oscillations.
It is to be noted that the shock oscillations mentioned above is especially when the entropy
at the inner sonic point is higher than that at the outer sonic point, not the signature of any instability.
In this work entropy of the pre and post-shock flow 
is not preselected. Flow starts with some set of energy and angular momentum and we calculate other physical parameters 
(say sound speed, flow velocity etc.), which are needed to calculate the entropy of the sonic points. After calculating entropy 
of the sonic points, we compare them. The number of critical points depend on the specific energy content in the flow as well. 
Since the number of critical points decides whether the flow will have shock waves or not. Thus sonic points, shocks and entropy 
all are inter related with the input parameters. If there are three sonic points, with higher entropy at inner sonic point, 
the flow tries to form a shock and the post-shock flow passes through the inner sonic point. However, since RH condition is 
not satisfied, this shock is not steady and oscillation starts. Numerical simulation by Ryu et al. (1997) observed 
that signature of oscillation. Similarly, if the cooling time scale in the post-shock region 
becomes comparable to the compressional heating time scale, then oscillation is a must, even if RH condition predicts 
a steady solution. This oscillation causes the QPO phenomenon.

From the above discussion on the behaviour of transonic flows, one arrives at the following 
conclusions: (a) A spectral fit of an outbursting object provides four physical parameters
including the time lag between peaks of the accretion rates of Keplerian and sub-Keplerian components. This, 
in turn, provides $\alpha_K$, i.e., the viscosity parameter in the Keplerian
component. (b) From the POS model, one obtains the shock location, i.e., the size of the Compton cloud, 
from the QPO frequencies, and from the transonic flow solution one can generate this shock 
when certain $\alpha_{SK}$, the viscosity parameter, is present in the sub-Keplerian component. (c)
Presence of the theoretically predicted value of a critical viscosity parameter $\alpha_{cr}$ was 
proven by numerical simulations and could be computed. For a complete consistency, we should see
$\alpha_{SK}<\alpha_{cr}<\alpha_{K}$ for each data.

In the present paper, we analyze several black hole candidates with outbursting properties
where both the rising and declining phases are included. We first find 
these viscosity parameters during the spectral state transitions independently 
for each observation for each object. When we compare them, an important 
property emerges. We find that in all these objects the state transitions from HS to HIMS 
take place at roughly specific values of $\alpha_{SK}$. At another $\alpha_{SK}$
HIMS to SIMS transition occurs. Furthermore, in each spectral fit using two component flow 
solution, we do find that this condition $\alpha_{SK}<\alpha_{cr}<\alpha_{K}$ is satisfied. We believe 
that these findings are very important in the context of  black hole  
accretion model and predictability in future outbursting sources.

The {\it paper} is organized in the following way: in the next Section, we discuss behavior of 
some of the transient sources during the outburst phase in details. In \S 3, we will present 
governing equations of modified RH shock conditions in presence of 
Compton cooling. In \S 4, we discuss the solution methodology in details.
In \S 5, we show how viscosity parameter varies in the disk on progressive 
days and what possible range of viscosity parameter triggers different spectral states of 
various candidates. We study the relation between the
QPO properties with viscosity parameter. Finally, 
in \S 6, we briefly discuss our results and make our concluding remarks.  

\section{Selected Candidates} 

We select five well known Galactic transient BHCs (GRO~J1655-40, H~1743-322, 
MAXI~J1543-564, MAXI~J1836-194, MAXI~J1659-152) to estimate the viscosity parameter on a daily basis. 
We use the rising phases of 2005 outburst of GRO~J1655-40, 2011 outburst of MAXI~J1543-564, 
2011 outburst of MAXI~J1836-194, 2010 outburst of MAXI~J1659-152 and the declining phase 
of 2010 outburst of H~1743-322 for our analysis. We mainly consider those cases when QPOs are 
present and the state transitions are prominently observed. For the calculation of viscosity parameter
we use RXTE/PCA data from 2.5 keV to 25.0 keV for all the candidates.
A summary of the basic properties of these objects are given in Table~1. 

\begin{table}  
\scriptsize
\vskip -0.4 cm
\centering
\caption{\label{table1} Details about the selected sources}
\begin{tabular}{|l|l|l|l|l|l|l|l|l|}
\hline
Source         &~ Mass          & Distance  &~ ~i     &~ ~Orbital      &Outburst &~Spectral&Discovery \\
&~$M_\odot$~   &~kpc~        &Degree  & Period (hr)    &Year &~$States^{*}$  &Year    \\
\hline 
GRO~J1655-40   &$7.02\pm0.22^{1}$ &$3.2\pm0.2^{2}$ &$69.5\pm0.1^1$& $62.920\pm0.003^3$ & 2005&HS, HIMS, SIMS &$1994^4$\\
H 1743-322     &$9.0 - 13.0^5$ & $8.5\pm0.8^6$ &$75\pm3^6$& -  &2010&HS, HIMS, SIMS  &$1977^{7, 8}$ \\
MAXI~J1543-564 &$12.6 - 14.0^9$  &$8.5^{10}$  & -      & -  &2011&HIMS, SIMS &$2011^{11a}$ \\
MAXI~J1836-194 &$7.5 - 11.0^{12}$   &$7\pm3^{13}$  &$4-15^{13}$ & $< 4.9^{13}$ &2011&HS, HIMS  &$2011^{11b}$ \\
MAXI~J1659-152 &$4.7 - 7.8^{14, 15}$ &$8.6\pm3.7^{16}$ &$70\pm10^{16}$& $2.414\pm0.005^{16}$ &2010&HS, HIMS, SIMS &$2010^{17}$ \\
\hline
\end{tabular}
\leftline {References: (1) Orosz \& Bailyn 1997, (2) Hjellming \& Rupen 1995, (3) van der Hooft et al. 2008, (4) Zhang et al 1994, (5) P\'etri 2008,}
\leftline {(6) Steiner et al. 2012, (7) Kaluzienski \& Holt 1977, (8) Doxsey et al. 1977, (9) Chatterjee et al. 2016, (10) Stiele et al. 2012,(11) Negoro et al. 2011ab,}
\leftline {(12) Jana et al. 2016, (13) Russell et al. 2014, (14) Molla et al. 2016, (15) Yamaoka et al. 2012, (16) Kuulkers et al. 2013, (17) Mangano et al. 2010.}
\leftline {*~ Observed spectral states}
\end{table} 
   
\section {Shock condition and viscosity parameter calculation}

In this Section, we present a prescription to obtain $\alpha_{SK}$ in the transonic
flow which includes cooling processes. We consider a thin, axisymmetric and radial 
flow in pseudo-Newtonian geometry. Specific angular momentum of the flow ($l$) produces a centrifugal
barrier close to the black hole since the centrifugal force $\sim l^2/r^3$ increases rapidly. 
Matter behind this barrier piles up and forms a shock (Chakrabarti 1989b, 1990). The post-shock region 
is hotter due to the conversion of kinetic energy into thermal energy and behaves like a 
Compton cloud. This region up-scatters soft photons from the Keplerian disk. Thus the
energy inside the CENBOL decreases and modifies the RH shock conditions. According to 
the nature of dissipation, three types of shocks were envisaged (Chakrabarti 1989b; Abramowicz \& Chakrabarti 1990). 
The conserved specific energy of the flow in pseudo-Newtonian geometry is given by (C89a),
$$
\varepsilon=\frac{\varv^2}{2}+\frac{1}{\gamma-1}{a}^2+\frac{l^{2}}{2r^{2}} - \frac{1}{2(r-1)},
\eqno{(1)}
$$
where $\gamma$, $\varv$, $l$ and $a$ are the adiabatic index, radial velocity of the flow, specific 
angular momentum of the flow and adiabatic sound speed defined by $\sqrt{\gamma P/\rho}$ respectively. 
Here $P$ is the isotropic pressure and $\rho$ is the gas density. 
As both the flows are relativistic hence have the same adiabatic index.
In this work we consider the adiabatic index 4/3 throughout. If we had magnetic fields or pair plasma, the index 
would be different. We do not consider these effects here.
In presence of cooling, conservation of energy equation of RH conditions take the form:
$$
\varepsilon_{+}=\varepsilon_{-} - \Delta \varepsilon,
\eqno{(2a)}
$$
where, $\varepsilon_{+}$, $\varepsilon_{-}$ and $\Delta \varepsilon$ are the energy of the post shock, pre shock flow and 
energy loss due to Comptonization respectively. Energy loss comes from observed spectral energy distribution,
given by,  $\sum_{i=\nu_{l}}^{\nu_{u}} \nu_i F_{Comp}(i),$ where, $\nu_{l}$, $\nu_{u}$ and $F_{Comp}$ are the lower
and upper limits of the frequency and Comptonized flux of the soft photons after it's scattering by the hot electron cloud 
inside the corona. The methodology of $\Delta \varepsilon$ calculation is given better in details in Mondal et al. (2015).
Baryon number conservation equation at the shock is,
$$
\dot{M_{+}}=\dot{M_{-}}.
\eqno{(2b)}
$$
Puffed up gas modifies the RH conditions so that pressure and density parameters become
important. The pressure balance condition is given by (C89a): 
$$
P_{+} + \rho_{+}\varv_{+}^2 = P_{-} + \rho_{-}\varv_{-}^2.
\eqno{(2c)}
$$
Here, ``-'' and ``+'' signs stand for quantities before and after shock.
After solving above equations, we get a condition which connects the mach number ($M$), shock constant 
($C_{shk}$, which connects pre- and post- shock mach numbers), cooling energy and pre shock sound speed ($a_{-}$), 
given by (C89a):

$$
A~M_{-}^4+B~M_{-}^2-C=0,
\eqno{(3)}
$$
where
$$
A=(\gamma-1)~C_{shk}-(3\gamma-1)^2,
$$
$$
B=2~C_{shk}\left[1-\frac{\Delta \varepsilon (\gamma-1)}{a_{-}^2}\right]-4(3\gamma-1),
$$
$$
C=4,
$$
where, $M_-$ is the pre-shock mach number of the flow. We follow the same mathematical procedure and solution 
technique as in Chakrabarti (1990), to find shock location for a given cooling.

Viscosity is an important physical parameter in standard 
thin accretion disk around black holes (SS73). Viscosity transports angular momentum 
of the inflowing matter outward and allows it to fall 
into the black holes. As the shock moves closer, the angular momentum must be adjusted by
viscosity so that the shock formation is allowed. For our calculation, we use 
the relations (for details see, Matsumoto et al. 1984; Chakrabarti 1990): 
$$
\dot{M_h}(l^{'}-l)=-X_s^{2}W_{r\phi},
\eqno{(4)}
$$
where, $\dot{M_h}$ and $W_{r\phi}=-\alpha_{SK} P$, are the sub-Keplerian mass accretion rate and viscous stress respectively, 
$\alpha_{SK}$ being (SS73) viscosity parameter of the sub-Keplerian component. 
We assume that the flow satisfies SS73 prescription at a given point of the flow so that viscous 
stress is proportional to the local pressure. Angular momentum variation from Eq. (4) can be written as (Chakrabarti 1990),
$$
\Delta l=\frac {\alpha_{SK} X_s a^{2}}{\varv},
\eqno{(5)}
$$
where, $\varv$ is the radial flow velocity, $X_s$ is the location of the shock in $r_g(=2GM/c^2$ 
where G, M and c are the gravitational constant, mass of the black hole and speed of light respectively) unit, 
and $\Delta l=(l^{'}-l)$ is the difference of specific angular momentum $(l)$ between
the specific angular momentum at some radius, from which flow starts compared to the radius where shock forms,
for a given cooling. Specific angular momentum at the radius where flow starts is $l^{'}$ and it is $l$ at the shock location, 
where observed QPO matches with theoretical. During the estimation of $\alpha_{SK}$ we are not using accretion rate directly, 
we are considering only the physical quantities such as the $a$, $\varv$ and $X_s$ etc., which come from basic input parameters 
for the sub-Keplerian component. The observed QPOs give us an handle on where the shock forms. That gives us the input 
parameters for the sub-Keplerian flow. The viscosity parameter is assumed to come from SS73 prescription which says that 
the viscous stress is proportional to the local pressure.

For the calculation of viscosity parameter of the Keplerian component ($\alpha_K$), we use the following approach as described in Nagarkoti \& Chakrabarti (2016b, hereafter NC16b). First, we calculate,
$$\frac{h}{r}= 2.4\times 10^{-3} \alpha_K^{\frac{-1}{10}}\dot{M_d}^{\frac{3}{20}}m^{\frac{-3}{8}}r^{\frac{1}{8}}r_*^{\frac{1}{8}}f^{\frac{3}{5}},
\eqno{(6)}
$$ 
(Frank et al. 2002). To calculate the vertical structure of the disk (Eq. 6), Frank et al. (2002) considered the 
local structure of thin disk, where most of the physical parameters (density, temperature, optical depth etc.) come from 
SS73 approximation. The temperature of the disk comes from an energy equation relating the energy flux in the vertical 
direction to the rate of generation of energy by viscous dissipation. We also assumed density and temperature in such a 
way that the Rosseland mean opacity is well approximated by Kramer's law. Eq. (6) shows that the disk is thin as long as 
SS73 disk solutions hold. Thus unless the disk solution breaks down, the disk can extend out to quite large radii, up to Roche lobe.
So these considerations that the disk remains thin till the piling up radius allow us to compute $\alpha_K$.
Then we use the relation: 
$$t_{visc} \approx \frac{r^2}{\alpha_K c_s h} \approx \frac{r^{1/2}(r-1)}{\alpha_K (\frac{h}{r})^2},
\eqno{(7)}
$$
(Pringle 1981). In these equations, 
$\alpha_K$, $c_s$, $r$ and $h$ stand for the Keplerian viscosity parameter, the adiabatic sound speed,
radius at the outer edge (in units of $r_*=\frac{Gm}{c^2}$), and the scale height of the accretion disk respectively,
$\dot{M_d}$ is peak accretion rate of the Keplerian disk in the units of $10^{17}erg~ s^{-1}$, $m$ 
is mass of the black hole in units of solar mass, 
$f=\bigg(1-\frac{1}{r^\frac{1}{2}}\bigg)^{\frac{1}{4}}$ and $t_{visc}$ stands for viscous timescale. 
Assuming both sub-Keplerian matter and disk matter are coming in from outside together during the 
outburst, the sub-Keplerian (low angular momentum flow) matter reaches the shock location in matter of seconds 
while Keplerian matter takes much longer ($\sim days$). Hence, $t_{visc}$ is taken as the time gap between 
the sub-Keplerian accretion rate peak and Keplerian accretion rate peak. In general, we choose outer edge of the disk such that the disk 
effective temperature $\sim10^4~K$, enough for hydrogen to become partially ionized. All other parameters are 
shown in Table~3 along with references. 
  
\section{Solution methodology}

In this manuscript, to calculate the $\alpha_{SK}$, we choose a set of specific energy ($\varepsilon$) and 
specific angular momentum ($l$) value of the flow. Due to the presence of cooling, R-H condition finds
a stable shock which produces the observed QPO frequency. This cooling (dissipated) energy on each day 
is calculated by integrating observed spectrum on that day. For a particular observation i.e., a set of 
cooling and QPO values, $l$ changes in each iteration and produces shock location. When theoretical and observed
QPOs match within $5$\% tolerance, we pick up that $l$ value and subtract from the original value by which flow
started, to get $\Delta l$ and viscosity parameter. To calculate QPO frequency using shock location we use
the relation $\nu_{QPO} = M_{BH}\times10^{-5} / [R~X_s(X_s-1)^{1/2}]$, where R ($\varv_-/\varv_+$) is shock 
compression ratio and $X_s$ is the location of the shock in $r_g$ unit (Debnath et al. 2014). 

We estimate $\alpha_{cr}$ using the following method (NC16b). We create a database using transonic flow theory 
the value of $X_s$ all over the parameter space. We then integrate inwards from the shock location to find 
the infall timescale, $t_{infall} =\int{dt}=\int {\frac{dr}{\varv}}$. The integration is carried out from the shock location to 
the inner sonic point. Here dr is the elemental radial distance in $r_g$ unit and 
$\varv$ is the infall speed obtained from the flow solution for the radial advection of dr.
The frequency of QPOs is found using the relation $\nu_{QPO} = \frac{1}{t_{QPO}} = \frac{1}{t_{infall}}$ (MSC96).
Then, we look at observed data (TCAF-fit, for GRO source we follow disk and halo rate from Table~5 of Debnath et al. 2008 ) and choose the range of viscosity parameter, $X_s$ and QPO which was 
observed. Then, we tally this theoretically possible values of $X_s$ and QPO frequencies with observational results. 
Essentially, the theory gives us all possible values from which we select what is relevant for the particular 
outburst of interest. The maximum $\alpha$ relevant for an outburst is the critical value of $\alpha_{cr}$.
To calculate, $\alpha_K$, we use Eq.(6-7) following NC16b procedure as explained in \S 3. 
  
\section{Results}

In this {\it Paper}, we study evolution of viscosity parameter in outbursting BHCs from theoretical 
consideration. We also estimate the range of viscosity parameter $\alpha_{SK}$ which triggers spectral state change. 
In Fig. 1a (black, cyan, magenta and green lines), we show the evolution of QPOs frequency 
in progressive days during the rising phase of different outbursts of the candidates. The opposite scenario 
is observed during the declining phase of the outburst of H~1743-322 (blue line). 
In Fig. 1b, we show the variation of viscosity parameter ($\alpha_{SK}$) with day for the candidates during the rising
and declining (blue line) phases of the outburst. We again observe that the viscosity parameter variation is
different for GRO source as compared to other sources. There is a significant variation in GRO J1655-44 which implies
that the rate of transport of angular momentum is not systematic. During first few days viscosity parameter is 
almost constant. After that, it increases rapidly and again increase slowly. In the last few days of the 
rising phase, viscosity parameter rises rapidly and shock reaches $\sim 13 r_g$ and produces 
$\sim 17.78 Hz$ QPOs. Green line shows variation of $\alpha_{SK}$ in progressive days for MAXI~J1543-564. 
We see that on the first day of the outburst, viscosity parameter starts with $\sim 0.01$.
This is reasonably high as compared to all other sources. This indicated that the shock is
not very far from the black hole and the spectral state is not hard. Similar signature is also observed in TCAF and POS model fitted results (Chatterjee et al. 2016). Detailed values of the sub-Keplerian viscosity parameter are shown in Table~2 for all the candidates. 

We calculate $\alpha_K$ for all the sources from the time-lag between the peaks of the accretion rates.  
Our calculated $\alpha_K$ is well above the $\alpha_{SK}$, which satisfies theoretical constraint 
(Chakrabarti, 1990; 1996) that $\alpha$
must be super-critical in order to form a Keplerian disk. Since the critical 
viscosity parameter changes with the input parameters, it is different from case to case. 
The values of $\alpha_K$ are $0.18, ~0.25, ~0.18, ~ 0.22$ and $~0.29$ for H~1743-322, 
GRO~J1655-40, MAXI~J1836-194, MAXI~J1659-152 and MAXI~J1543-564 respectively.          
The maximum value of viscosity parameter for which standing shocks are still possible (i.e., for
$\alpha_{cr}$) is found for all these candidates. 

In Fig. 2, we show the variation of $\alpha_{SK}$ with shock location in logarithmic scale. Different curves 
(with the same colour code) show variation for different candidates during their outburst 
time. We see that all the variations follow the same profile. It is clear however that the viscosity in the 
sub-Keplerian component must rise rapidly in order to achieve 
spectral state transitions. Eventually, when they catch up with that of the Keplerian 
component, the soft state occurs. So an outburst may be incomplete (i.e. the soft state is not reached)
if $\alpha_{SK}$ is not sufficient.

\begin{table}
\scriptsize
\centering
\caption{\label{table2} Theoretically calculated parameters}
\begin{tabular}{|l|l|l|l|l|l|ccccc|}
\hline
Source & Obs. Date &$\Delta \varepsilon$ & $X_s$ &$\alpha_{SK}$ & QPOs\\
\hline
H 1743-322  &55455.44&0.176E-3&38.48&\textbf{0.681E-1}&6.417 \\
 (Dec.)     &55456.68&0.165E-3&63.92&0.329E-1&3.276 \\
            &55457.12&0.154E-3&77.01&0.246E-1&2.569 \\
            &55458.64&0.133E-3&105.10&0.146E-1&1.761 \\
            &55459.68&0.119E-3&148.21&0.761E-2&1.172 \\
            &55462.56&0.093E-3&227.92&0.292E-2&0.741 \\
            &55465.12&0.065E-3&1357.5&0.477E-5&0.102 \\
            &55467.52&0.048E-3&1327.0&0.483E-5&0.149 \\
            &55469.01&0.038E-3&1625.1&0.574E-5&0.079 \\
\hline
GRO J1655-40&53426.04 &0.898E-5&1506.8&0.532E-5 & 0.082\\
 (Ris)      &53427.15 &0.111E-4&881.16&0.143E-3 & 0.106\\
            &53428.13 &0.120E-4&755.93&0.261E-3 & 0.116\\
            &53428.85 &0.118E-4&695.99&0.350E-3 & 0.122\\
            &53431.61 &0.114E-4&636.01&0.472E-3 & 0.129\\
            &53432.79 &0.139E-4&462.41&0.119E-2 & 0.163\\
            &53433.90 &0.193E-4&295.03&0.347E-2 & 0.244\\
            &53434.69 &0.223E-4&229.25&0.579E-2 & 0.316\\
            &53435.61 &0.302E-4&192.12&0.807E-2 & 0.382\\
            &53436.15 &0.351E-4&178.53&0.921E-2 & 0.416\\
            &53436.39 &0.351E-4&160.18&0.111E-1 & 0.471\\
            &53437.07 &0.384E-4&148.95&0.126E-1 & 0.513\\
            &53438.75 &0.461E-4&82.08 &0.153E-1& 1.349\\
            &53439.10 &0.866E-4&73.41 &0.188E-1& 1.526\\
            &53439.74 &0.993E-4&52.31 &0.334E-1& 2.313\\
            &53440.73 &0.120E-3&24.44 &0.444E-1& 6.522\\
            &53441.51 &0.393E-3&13.27 &\textbf{0.547E-1}& 17.78\\
\hline
MAXI J1543-564&55691.09&0.541E-4&155.70&0.823E-2&1.046\\
 (Ris.)       &55692.09&0.562E-4&102.04&0.176E-1&1.753\\
              &55693.09&0.673E-4&67.43&0.339E-1&2.978\\
              &55694.10&0.785E-4&50.69&0.511E-1&4.327\\
              &55694.89&0.852E-4&41.08&\textbf{0.679E-1}&5.700\\
\hline
MAXI J1836-194&55804.52&0.272E-4&567.25&0.912E-5 &0.476\\
 (Ris.)       &55806.51&0.359E-4&254.27&0.875E-3 &0.895\\
              &55812.57&0.428E-4&141.59&0.459E-2 &1.530\\
              &55819.20&0.350E-4&51.25& 0.331E-1 &4.876\\
              &55820.40&0.327E-4&49.33& \textbf{0.351E-1} &5.175\\
\hline
MAXI J1659-152&55467.19&0.398E-3&281.96&0.369E-3 &1.607 \\
 (Ris.)       &55468.09&0.365E-3&190.43&0.146E-2 &2.278 \\
              &55469.09&0.429E-3&153.79&0.277E-2 &2.723 \\
              &55470.26&0.422E-3&153.68&0.277E-2 &2.749 \\
              &55471.51&0.397E-3&139.42&0.356E-2 &3.028 \\
              &55472.07&0.399E-3&127.72&0.445E-2 &3.329 \\
              &55473.47&0.468E-3&94.81 &0.890E-2 &4.709 \\
              &55475.43&0.463E-3&74.36 &0.146E-1 &6.108 \\
              &55476.67&0.519E-3&66.47 &\textbf{0.181E-1} &6.981 \\
\hline
\end{tabular}
\leftline {Observation dates are in MJD (Modified Julian Day)}
\leftline {Only frequency of the primary dominating QPOs} 
\leftline {are mentioned.}
\end{table}

\begin{table} 
\scriptsize	
\vskip -0.4 cm
\centering
\caption{\label{table3} Parameters and calculated $\alpha_K$ values}
\begin{tabular}{|l|l|l|l|l|l|l|}
\hline
Source         &~ $m$          &~$\dot{m_d}$       &~$t_{\it visc}$ &~$\alpha_K$  &$~\alpha_{cr}$ &~$\alpha_{SK}$ \\
               &~($M_\odot $)~ &~($\dot{M}_{Edd})$~&~(Days)~        &             &               &       \\
\hline 
H 1743-322     &11.4  &3.600  &13.6   &0.18  &0.13 &0.068   \\
GRO~J1655-40   &7.02  &1.740  &7.3    &0.25  &0.16 &0.018   \\
MAXI~J1543-564 &13.2  &0.656  &7.7    &0.29  &0.14 &0.055   \\
MAXI~J1836-194 &8.0   &2.395  &10.0   &0.18  &0.11 &0.068   \\
MAXI~J1659-152 &6.0   &1.328  &7.2    &0.22  &0.10 &0.035   \\
\hline
\end{tabular}
\leftline {References: Mondal et al. 2014, Orosz \& Bailyn 1997, Debnath et al. 2008}
\leftline {Chatterjee et al. 2016, Jana et al. 2016, Molla et al. 2016}
\end{table} 

\subsection{Spectral state transitions triggered by viscosity}

Most of the BHCs show a complete cycle of spectral states during their outburst time. Hardness-Intensity 
diagram (Fender et al. 2004; Belloni et al. 2005) shows that hysteresis behavior and spectral states using
truncated disk model and QPO information. In TCAF solution, this is due to the fact that Keplerian disk is 
easier to form in the rising state, but after the withdrawal of viscosity, it does not disappear 
in the same time scale (Roy \& Chakrabarti 2017). In the present paper, we assume that the
main cause of high accretion rate close to the inner edge is because of enhancement of 
viscosity and find that the viscosity parameter must achieve a minimum value 
to initiate a certain spectral state. In Fig. 1a, we show the variation of 
QPO frequencies with day both theoretically calculated and the observed values. If the viscosity 
parameter varies with day as shown in Fig. 1b, then theoretical and observed QPOs 
will vary together in the same way. Thus viscosity parameter of the 
outbursting candidates should change (without the pattern shown) rather than be a constant value.   

\noindent{\bf H~1743-322:} Mondal et al. (2014) showed from TCAF fit that in MJD 55462.62,
during the declining phase of the outburst, accretion rate ratio (ARR) is high and spectral state goes 
from hard-intermediate state to hard state. Here we see that viscosity parameter is reduced and the shock moves 
far away from the source. Debnath et al. (2013) studied evolution of the QPO frequencies during the rising and 
the declining phases of two successive outbursts (2010 and 2011) with POS model. 
This candidate satisfies resonance condition during its 2010 rising phase of the outburst (Chakrabarti et al. 2015). 

\noindent{\bf GRO~J1655-40:} At the initial phase of the outburst, the QPO frequencies
were very low. After MJD 53437.07, shock moves faster, thus the Compton cloud 
collapses rapidly and QPO frequency increases (see, Table~2). Sub-Keplerian viscosity parameter also
started increasing after this date. We observe that after this date, transition to
an intermediate state takes place. Day-wise evolution of QPO frequencies during rising 
and declining phase are fitted with POS model (Chakrabarti et al. 2005; 
Debnath et al. 2008) to find instantaneous location, velocity, etc. of the shock. 
A monotonic rise in QPO frequency from $82$~mHz to $17.78$~Hz in this phase was observed, 
with an inward movement of the shock wave due to more and more supply of the highly 
viscous Keplerian matter as day progressed. Simultaneous non-harmonic QPO and spectral
behavior using HID is also studied by Motta et al. (2012).  

\noindent{\bf MAXI~J1836-194:} 
During this outburst, a clear dominance of the low-angular momentum sub-Keplerian component over 
the high viscous Keplerian flow was observed (see, Jana et al. 2016). 
This could be due to small accretion disk as because of shorter orbital period of the binary system or 
it may be immersed in the wind or the excretion disk of the companion of high massive Be star. 
Authors also estimated the viscous time scale around $10$~days, peak time differences 
between Keplerian and sub-Keplerian accretion rates, observed during the outburst. 
On MJD 55812.57, the shock moves in significantly due to the effects of cooling,
which produces hard intermediate state. This is an indication of increasing viscosity parameter.
On the highest QPO frequency day (MJD 55820.40), $\alpha_{SK}$ is the highest thereby 
ushering the transition to SIMS. 

\noindent{\bf MAXI~J1659-152:} Here the behavior of cooling is quite different. This may be due to the 
presence of strong wind accretion (Debnath et al. 2015) and the Keplerian disk is always 
inside the halo. 
On MJD~55469 and MJD~55470, cooling is constant thus the viscosity parameter and QPOs 
are constant. On MJD 55472.07, viscosity increased significantly and also QPO frequency started increasing. 
The soft spectral state was absent. This may be due to high accretion rate from low-angular momentum 
component over the highly viscous Keplerian component. This candidate showed
X-ray emission during quiescence, which indicates that there is always an accretion disk present 
in the system. Kuulkers et al. (2013) proposed that the outburst of the system is most
likely due to some disk instability mechanism. Mass loss event from a companion filling its Roche lobe is 
unlikely to be the cause. 
The high flow of matter in the outburst is the result of convective instability at the radius where matter was piling up. 
That raised the viscosity and caused the formation of the Keplerian disk. The viscosity of the sub-Keplerian flow also changed. 
But the triggering is done far away at the convective radius. 

The mass accretion rate variability has been observed by many authors (Kuulkers et al. 2013; Kalamkar et al. 2015; Uttley \& Klein-Wolt 2015) 
for black hole candidates considered in this paper. We also see from the fitted data (Debnath et al. 2014; 
Mondal et al. 2014; Jana et al. 2016; Molla et al. 2016) that the accretion rates of the observed candidates change daily basis
during the outburst. Thus viscosity changes accretion rate also clear from SS73 as well as from our transonic solution.
  
\subsection{Estimated range of triggering}

In Chakrabarti (1990, 1996), it was shown that viscosity parameter decides the topology of the flow and particularly
above $\alpha_{cr}$, the topology changes from a sub-Keplerian flow 
topology with or without a shock to a standard Keplerian disk type topology. 
Since spectral states depend on the radiation contributed by these two types of flows, the viscosity parameter 
plays a crucial role. On the basis of the present analysis (Table~2) we find that if the viscosity parameter in 
the sub-Keplerian component is  $< 0.008$, the source will be in pure HS. In between $0.008-0.035$ of the 
$\alpha_{SK}$, the source will be in HIMS. For the value $0.035-0.1$, it will show SIMS and if the viscosity parameter 
value is $>0.1$, source will be in soft state. 
In Fig. 2, we show the probable limits of state change in different shaded regions. From Table~2,
we see that the maximum value of the viscosity parameter ($\alpha_{SK}$) is $< 0.1$ for these sources, 
when they are in SIMS. This estimation is made on the basis of selected candidates and their date of state change. 
As our analysis covers most of the black holes mass range, we believe this analysis also covers other sources which we have not 
analyzed yet. This result is within the range of viscosity  parameter limit calculated by advection scenario and MRI simulations. 
It has been shown by NC16b that in Keplerian components of the MAXI~J1836-194 and MAXI~J1659-152 sources, 
the $\alpha_{K}$ is $0.18$ and $0.22$ respectively, which is more than the above critical values, which is 
consistent. In Giri \& Chakrabarti (2013) it was shown that the super-critical viscosity 
on the equatorial plane clearly leads to the segregation of the original inflow into TCAF of CT95. So the
theoretical conjecture of two component formation due to variation of viscosity and their emission of 
two components of radiation can be considered to be close to reality while considering black hole accretion process.
    
\section{Discussions and Concluding Remarks}

It is believed that the viscosity is a crucial physical parameter to trigger the outbursts
in transient sources. In this paper we estimate the range of $\alpha_{SK}$ parameters for different 
spectral states using TCAF solution. During the outburst, a BHC shows QPOs.
These QPOs are very stable features in BHCs and observed day after day. 
If we actually compute the QPOs from theoretical model using radiated energy loss from a 
self-consistent transonic solution, we see that the shock drifts towards the black hole and thus QPO
frequency rises in the rising phase. 
The shock may be oscillating stably around a mean location (see MSC96; Garain et al. 2014). 
However, in presence of changing cooling factor (such as when the accretion rate changes in ourbursts), the mean location of the 
shock also shifts (see, Mondal et al. 2015). Thus the oscillating shock can also propagate towards the black hole, i.e. 
mean location will be drifting inwards in the rising state. The shock will drift outwards in the declining state.
The modified shock location and QPO frequencies are determined by RH conditions.
It is already discussed that according to TCAF, QPOs are due to shock oscillations. 
Oscillation of shock causes the Compton cloud size to oscillate and thus the reprocessed photons number density also fluctuates.
Since shocks can form, and indeed are stronger, in the absence of viscosity, QPOs will always form whenever shocks
oscillate even in the absence of viscosity. However viscosity plays a role indirectly. In presence of viscosity
the shocks shift their positions and thus QPO frequency changes. Now the question arises, does QPO ceases to exist when 
$\alpha_{SK}$ goes to zero in this model? If so, what is the physical reason for that? The answer is, QPO exists even 
with $\alpha_{SK}$ is zero. Only cooling process should be present. QPO directly does not depend on $\alpha_{SK}$. 
It depends, but via shock location which depends on $\alpha_{SK}$. When the oscillation
is more or less sinusoidal, we will have a single peak. However, normally the shock oscillation amplitude is not sinusoidal
(see, MSC96). In that case, Fourier transform does not yield a single peak and higher harmonics are
also possible. Since shock is not very thin, its oscillation at various distances is not at the same frequency and
thus the QPO does not look like a delta function and it has a width.

Recently, Le et al. (2016) studied the accretion properties from global instability of the disk. For that authors gave a perturbation
in shock velocity and study that the QPOs of harmonics 2:3 and 3:5 are possible, when the shock is located near the horizon and the
fundamental mode and overtones are suppressed by some nonlinear effects. In their simulation authors fixed the inner sonic point
so that perturbation given in shock will be zero when it reaches the inner sonic point. The harmonics estimate from the simulation is
also observed by several authors (Cui et al. 1999, Remillard et al. 2002). According to Chakrabarti et al. (2006), 2:3 ratio 
QPOs are observed due to non-axisymmetric effects when the shock switches between the two armed and the three-armed spirals. 
Observational evidences show that 2:3 ratio mainly observed in high frequencies, when shock is closer to the black hole and 
viscosity is also higher mainly in the soft spectral states. At that state accretion rate is higher and axisymmetric shock is 
pushed inside due to ram pressure, it breaks partially and causes the spiral structures to form.
In the present work we are not considering spiral shocks to produce harmonics, which will be reported elsewhere. 

Our observed result gives a range of viscosity parameter for the transient sources, which is $< 0.1$  to achieve
hard and intermediate states. Our result agrees with the recent analytical solutions (NC16ab).
Furthermore, from Table~3, it is clear that $\alpha_{SK}<\alpha_{cr}<\alpha_{K}$ as expected from theoretical considerations.
The viscosity parameter variation with day (Fig.~1b) shows that when the outburst starts (HS)
the shock is very far from the black hole and the viscosity parameter is very low (except, MAXI~J1543-564 which 
started in HIMS). 
As the viscosity parameter increases, shock moves in and the spectral state goes to intermediate states. 
During the declining phase of H~1743-322, we see the same trend in viscosity parameter, namely, 
as the day progresses (considering the starting of the declining phase as the 
first day) viscosity parameter decreases and the shock moves further away. 
Thus viscosity parameter does rise and fall and causes the cycle of
the outburst. We discussed possible limits of viscosity parameters to trigger 
from one spectral state to other. The result is widely applicable since we see that
the masses vary by a factor of two. Thus the evolution of the spectral states, the outbursts and the QPO frequencies
are self-consistently understandable from our analysis. This result not only couples the changes in flow topologies
with the radiation properties, it also establishes our firm belief in the theoretical paradigm
while understanding the observed data. Some other sources e. g., Cyg~X-1, GX~339-4, GRS~1915+105 etc. will also be 
studied in future to find how viscosity affects their flow topologies. 
    
\section*{Acknowledgments}
We are thankful to the anonymous referee for making valuable comments and suggestions.
SM acknowledges FONDECYT post-doctoral grand (\# 3160350) sponsored by Govt. of Chile, for the work. SM also
acknowledges MoES sponsored (by Govt. of India) post-doctoral research fellowship during which this work was started.
SN acknowledges the support of a fellowship from Abdus Salam International Centre for Theoretical Physics, Italy
granted to ICSP. This research has made use of data and software provided by the HEASARC.

\begin{figure} 
\vspace {0.5cm}
\centering{
\includegraphics[scale=0.6,angle=0,width=8truecm]{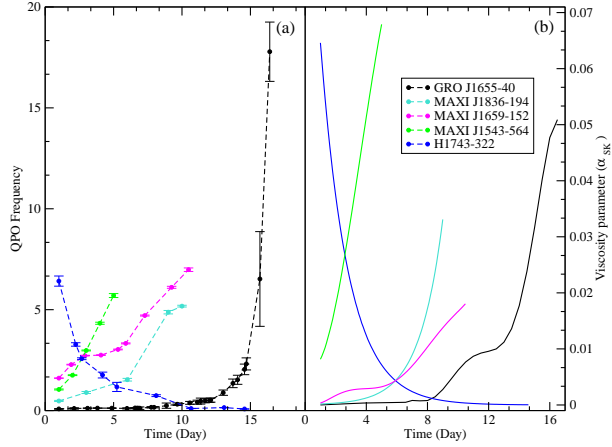}}
\caption{Variation of (a) QPO frequency with progressive days during different outbursts of the candidates, 
obtained from observation (circles) and analytical solution (dashed lines), and (b) viscosity parameter ($\alpha_{SK}$) with time (in day) during the outburst of the BHCs are shown.}
\label{fig1}
\end{figure} 

\begin{figure} 
\vspace {0.5cm}
\centering{
\includegraphics[scale=0.6,angle=0,width=8truecm]{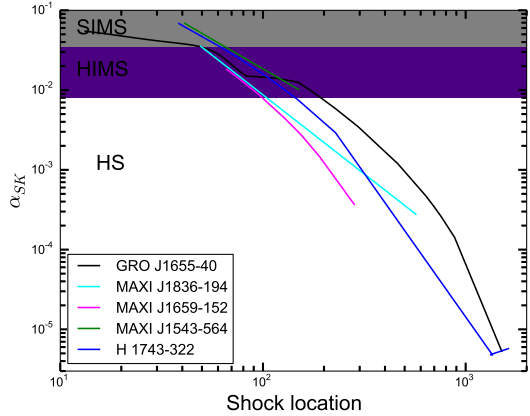}}
\caption{Variations of $\alpha_{SK}$ with shock location are shown in logarithmic scale. Different colours are 
for different sources. Different bands of $\alpha_{SK}$ are based on spectral transition dates for different candidates.
Here viscosity is plotted only for the sub-Keplerian component of the accretion flow.}
\label{fig2}
\end{figure}

%%%%%%%%%%%%%%%
\end{document}